# GENERATION OF SPECTRALLY-EFFICIENT SUPERCHANNEL USING OPTICAL FREQUENCY COMB REFERENCING AND ACTIVE DEMULTIPLEXING

Yi Lin[1], Seán Ó Duill[1,2*], Frank Smyth[2], Skip Williams[3], Anatoliy Savchenkov[3], Liam P. Barry[1]

[1]Radio and Optics research Laboratory, School of Electronic Engineering, Dublin City University, Dublin 9, Ireland
[2]Pilot Photonics Ltd., Invent Building, Dublin City University, Dublin 9, Ireland
[3]OEwaves Inc., 465 N Halstead St #140, Pasadena, CA 91107, USA.
*sean.oduill@dcu.ie

## Abstract

We generate multiple optical carriers with ultra-low phase noise, over a useable bandwidth of 160 GHz, from an externally injected gain-switched comb source with exceptional low linewidth below 10 Hz. We show successful transmission of 17 demultiplexed channels using 64-quadrature amplitude modulation signals at 5 GBaud.

## 1 Introduction

A limit on the single wavelength capacity for optical fiber systems is approaching [1] with the capacity per wavelength likely to maximise below 1 Tbit/s per wavelength. The requirement to switch channels with capacities >1 Tbit/s will require all of that data to be carried on two or more wavelength channels. Such multi-wavelength channels are termed superchannels [2]. The concept of a superchannel transmitter is shown in Fig. 1(a), and here we present a superchannel light source scheme in which free-running optical carriers are referenced to an ultra-low phase noise comb providing the ability to carry signals with 64-quadrature amplitude modulation (QAM) format. The architecture lends itself well to photonic integration and enables a single linewidth reducing element to be shared among many optical carriers in a miniature and cost-effective fashion.

Optical frequency combs [3] naturally lend themselves as candidates as multi-wavelength optical sources for superchannels, and have the additional benefit that the wavelength of the carriers do not drift independently thus offering the possibility to minimise the spectral guardband needed between the channels as well as accurately predicting/cancelling nonlinear transmission impairments due to cross-phase modulation [3]. There are some challenges in deploying optical frequency combs: firstly each comb line needs to demultiplexed separately to be modulated with data; secondly, the optical carriers of the comb need to have low phase noise in order to be able to carry modulation formats with high order cardinality of constellation points; and finally each comb line tends to have low power and typically needs to be amplified to allow for modulation and transmission.

In this submission we demonstrate a source suitable for superchannel transmission with high channel power per-line (10 mW) and independently modulated comb-referenced carriers that have exceptionally low linewidth. The superchannel is formed by actively demultiplexing lines from the optical frequency comb using injection-locking. Schematics of the optical frequency comb generation and active demultiplexing are shown in Fig. 1(b) and Fig. 1(c) respectively. The key enabler is the use of a single wavelength master laser within the optical frequency comb generator. The master laser is an OEwaves Hi-Q OE4030 laser based on stabilising the laser to a high quality factor microring resonator [4]. The linewidth of this master laser is exceptionally low in the sub-Hz range [5] and this ultra-low linewidth is transferred to all of the comb lines during the comb formation process. The comb is formed by gain-switching a distributed feedback (DFB) laser and more details are given in [6]. For this submission we generate an optical frequency comb with comb line spacing (or free spectral range (FSR)) of 10 GHz though the FSR is easily tunable for combs from gain-switched lasers [6]. We initially verify that the low-linewidth properties of the master laser are transferred to all of the comb lines. We then use this comb and demonstrate that it is a suitable source as an optical superchannel transmitter, by actively demultiplexing each comb line through injection locking of DFB lasers. We show that for seventeen of the demultiplexed comb lines from our comb, that transmission of 64-QAM at 5 Gbaud is possible with the received bit error rate (BER) falling below the 7% forward error correction (FEC) limit except for two channels that fall between the 7% and 20% FEC limits. From a superchannel perspective: the ability to transmit using seventeen individual channels implies that any configuration of wavelengths and channel spacing is possible over a 160 GHz useable superchannel bandwidth using this comb-referenced superchannel source. We also note that transmitting 64-QAM is a *record* highest cardinality of

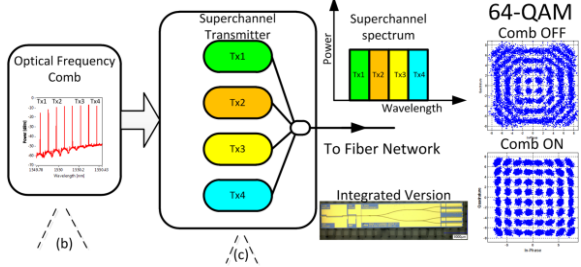
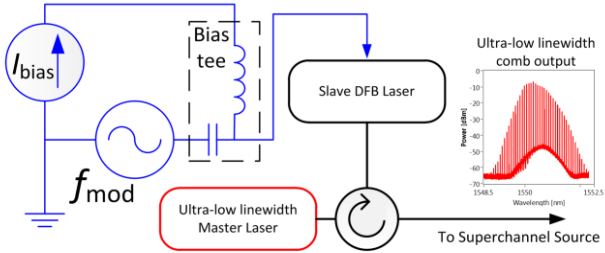
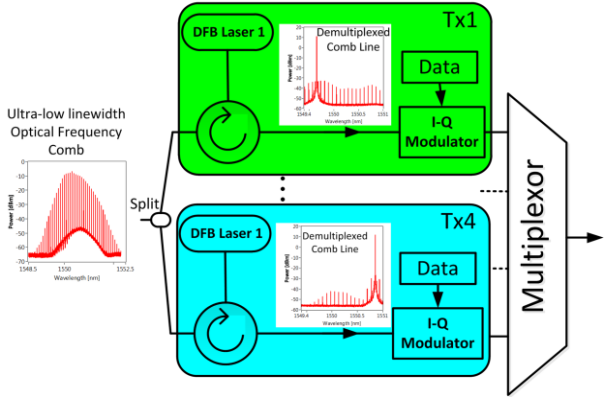
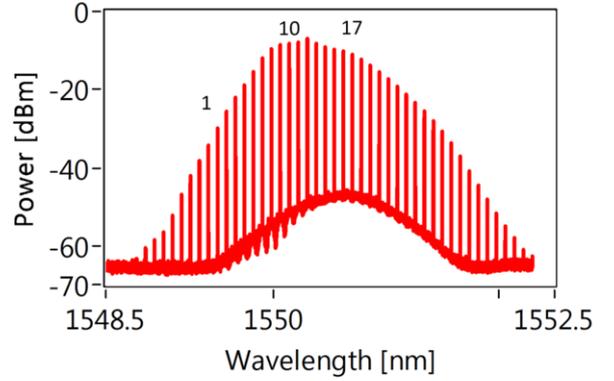
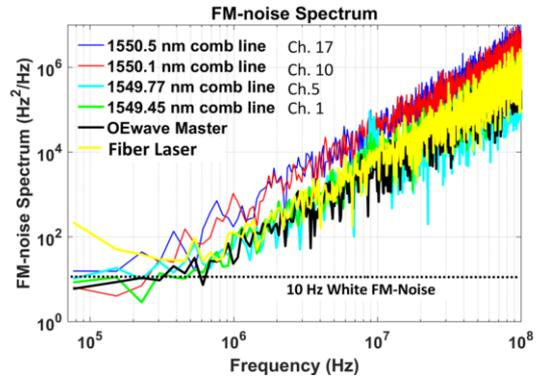

Fig. 2 (a) Spectrum of the ultra-low linewidth comb at 10 GHz FSR. The channel numbers of the respective comb lines are indicated (b) FM-noise spectra of the OEWaves lasers and comb source. Note the FM-Noise approaches 10 Hz and is better than a fiber laser.

Fig. 1 (a) Concept of a multi-wavelength transmitter that is referenced to a comb source. As shown, the ultra-low linewidth comb is necessary to transmit 64-QAM format. An integrated version of the comb and demultiplexor chip is also shown as inset. (b) Operational schematic of the gain-switched comb source, showing the placement of the ultra-low linewidth laser as the master laser. The actual generated comb is inset. (c) Injection-locking a bank of DFB lasers for use as the multiple carriers within a superchannel transmitter. The actual spectra of the demultiplexed comb lines are inset.

constellation points in a transmission format for optical transmission systems based on optical frequency combs from gain-switched lasers. These results demonstrate the feasibility to create a highly-integrateable optical superchannel source requiring only one ultra-low-linewidth laser source.
.
## 2 Low-phase noise comb generation and demultiplexing

The ultra-low linewidth optical frequency comb, as shown in Fig. 1(b) is generated via gain-switching the slave laser and injection locking using the ultra-low linewidth master laser. The linewidth properties of the master laser are transferred to each line in the comb. The slave laser was gain-switched at 10 GHz, therefore the comb line spacing (or FSR) is 10 GHz. The comb spectrum is shown in Fig. 2(a). The FM-noise spectra of the master laser and the filtered lines from the comb source were measured using delayed-self-heterodyne method [7]. The FM-noise results are shown in Fig. 2(b). The value of the FM-noise of the master laser descends below 10 Hz, and the increase in the FM-noise at higher frequencies is the noise floor of our measurement technique. Likewise the FM-noise of the comb lines approach the same value as the master laser. This is lowest recorded FM-noise for a gain-switched comb. The FM-noise of a fiber laser is also shown for comparison, and the ultra-low linewidth comb outperforms the fiber laser.

The comb is actively demultiplexed by injection-locking the DFB laser in the superchannel transmitter to one of the comb lines. More details about the comb demultiplexor can be found in [8]. The DFB laser in the comb demultiplexor injection-locks to the spectrally-closest comb line; therefore a different comb line can be demultiplexed by appropriately tuning the DFB laser. The spectra of the demultiplexed comb lines using a DFB laser are shown in Fig. 3, note that the exact same comb was used throughout and also that the suppression of the unwanted comb lines exceeds 40 dB.

## 3 Transmission Results

Now that the comb-referenced superchannel transmitter has been described, we proceed to show results of the 64-QAM

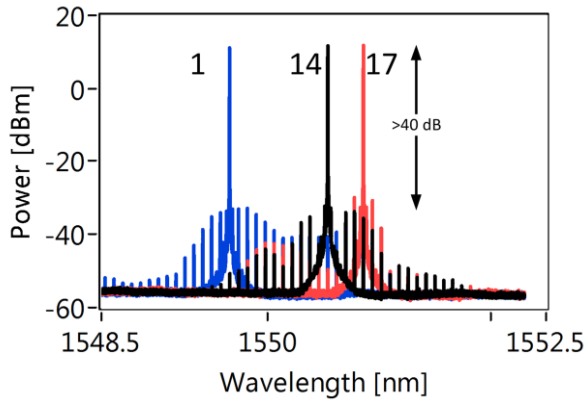

Fig. 3  Spectra of demultiplexed comb lines. Note the suppression of the residual comb lines exceeds 40 dB. Channels numbers 1, 14 and 17 are shown.

transmission. A schematic of the experimental setup and the offline-DSP is shown in Fig. 4. One line from the ultra-low linewidth optical frequency comb is selected by injection-locking a DFB laser in the superchannel transmitter. Different lines from the low-linewidth comb can be demultiplexed by tuning the DFB laser so that the free running wavelength of the DFB is within the locking range of the required comb line. The light from the DFB laser is modulated using 64-QAM format at 5 Gbaud. The modulated signal is then transmitted over 25 km of standard single mode fiber before detection at the coherent receiver. The main item in the offline-DSP for carrier phase recovery is the decision-directed phase locked loop (DD-PLL) [9]. The BER results are shown for demultiplexing of each comb line (channel number) in Fig. 5(a), the BER for each channel is below the 20% FEC limit and all but two channels below the 7% FEC limit. and therefore each channel could be used as a transmitter in a superchannel source. The total number of comb lines that showed successful transmission is 17, therefore the total spectral span of the superchannel is 160 GHz (10GHz FSR × (17-1)). This result implies that is possible to create any superchannel i.e. any possible baudrate, any number of channels provided that the selected carriers are within this 160 GHz useable spectral range, with no consideration is needed for spectral guardbands because of the comb-referencing..

To unequivocally show the impact of referencing to the low-linewidth comb, we present the received constellations with and without using the injected comb in Fig. 5(b). Clearly the DFB lasers are not be able to transmit the 64-QAM encoded data unless injection-locked to a high-quality, ultra-low phase noise comb line.

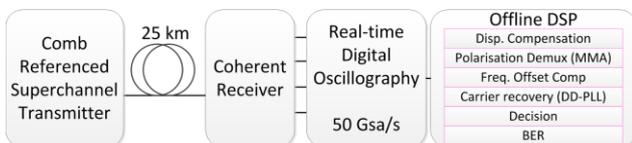

Fig. 4 Experimental setup for 64-QAM transmission using the superchannel transmitter source.

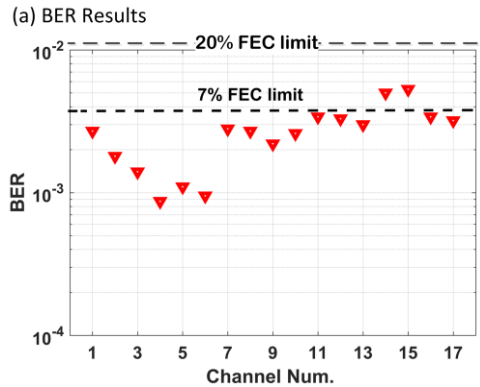

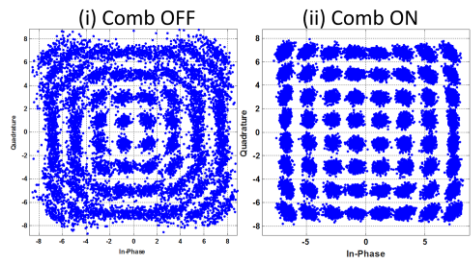

Fig. 5  (a) BER results for each demultiplexed comb after 25 km (▼). (b) Received constellations for the case (i) without referencing to the ultra-low linewidth comb and (ii) with referencing to the ultra-low-linewidth comb.

## 4  Conclusion

We have successfully shown the ability to transfer the ultra-low phase noise of a single laser to multiple optical transmitters using optical frequency comb generation and active demultiplexing. The system allowed us to transmit data encoded on 64-QAM format, the highest reported constellation cardinality for a gain-switched laser comb system. The scheme showed successful transmission of all possible channels at 10 GHz granularity over a 160 GHz bandwidth, which would be more than sufficient bandwidth to create 1.6 Tbit/s superchannels. For this submission we chose 10 GHz FSR and this sets the channel granularity, an important property of the gain-switched comb source is that the channel granularity can be easily and continuously tuned by varying the frequency of the gain-switching signal without changing the overall useable bandwidth, thus allowing for broad range of superchannel configurations. Work is underway to integrate (see inset of Fig. 1(a)) the comb and demultiplexor laser system onto a single chip [10] paving the way for flexible and powerful comb-based superchannel transceivers.

## 5  Acknowledgements

Dublin City University acknowledges Science Foundation Ireland: IPIC (12/RC/2276), CONNECT 13/RC/2077. Pilot Photonics acknowledges funding from the Enterprise Ireland DTIF Programme under grant No. 164576/7.

# 6 References

[1] Dr. Sorin Tibuleac "Coherent modulation in next-generation optical networks," Lightwave online https://www.lightwaveonline.com/network-design/high-speed-networks/article/16676197/coherent-modulation-in-nextgeneration-optical-networks (accessed Sept. 2019)

[2] P. J. Winzer, D. T. Neilson, and A. R. Chraplyvy, "Fiber-optic transmission and networking: the previous 20 and the next 20 years [Invited]," OSA Opt. Expr. 24190-24239, 2018

[3] N. Alic, E. Myslivets, E. Temprana, B. P.-P. Kuo, and S. Radic, "Nonlinearity Cancellation in Fiber Optic Links Based on Frequency Referenced Carriers," IEEE/OSA J. of Lightwav. Technol. 32, No. 15, p.p. 2690-2698, 2014

[4] W. Liang, V. S. Ilchenko, A. A. Savchenkov, A. B. Matsko, D. Seidel, and L. Maleki," Whispering-gallery-mode-resonator-based ultranarrow linewidth external-cavity semiconductor laser," OSA Opt. Lett.35, No. 16, p.p. 2822-2824, 2010.

[5] S. Williams, V. Savchenkov, "OEWaves Laser test Datasheet OE4023/4026/4028/4030 TDS10003 Rev A ", Private commun.

[6] P. M. Anandarajah, R. Maher, Y. Q. Xu, S. Latkowski, J. O'Carroll, S. G. Murdoch, R. Phelan, J. O'Gorman, and L. P. Barry, "Generation of Coherent Multicarrier Signals by Gain Switching of Discrete Mode Lasers," IEEE Photon. Journ. **3**, No. 1, pp 112 – 122, 2011.

[7] T. N. Huynh, L. Nguyen, and L. P. Barry," Phase Noise Characterization of SGDBR Lasers Using Phase Modulation Detection Method With Delayed Self-Heterodyne Measurements," IEEE/OSA J. of Lightwav. Technol. 31, No. 8, p.p. 1300-1308, 2013.

[8] R. Zhou, T. Shao, M. D. G. Pascual, F. Smyth, and L. P. Barry, "Injection locked wavelength de-multiplexer for optical comb-based |Nyquist WDM system," *IEEE Photon. Technol. Lett.* **27**, No. 24, 2595 – 2598, 2015. Available: http://dx.doi.org/10.1109/LPT.2015.2478791

[9] T. N. Huynh, A. T. Nguyen, W.-C. Ng, L. Nguyen, L. A. Rusch, and L. P. Barry," BER Performance of Coherent Optical Communications Systems Employing Monolithic Tunable Lasers With Excess Phase Noise," IEEE/OSA J. of Lightwave. Technol. 32, No. 10, p.p. 1973 – 1980, 2014.

[10] M. D. G. Pascual, J. Braddell, F. Smyth, and L. P. Barry, "Monolithically Integrated 1x4 Comb De-multiplexer Based on Injection Locking," presented at 18th European Conf. on Integrated Opt., ECIO-P37, Warsaw (Poland), 2016.